\documentclass{article}

\usepackage{arxiv}

\usepackage[utf8]{inputenc} 
\usepackage[T1]{fontenc}    
\usepackage{hyperref}       
\usepackage{url}            
\usepackage{booktabs}       
\usepackage{amsfonts}       
\usepackage{nicefrac}       
\usepackage{microtype}      
\usepackage{lipsum}		
\usepackage{graphicx}
\usepackage{natbib}
\usepackage{doi}

\title{Is Cambodia the World's Largest Cashew Producer?}

\author{
    Veasna Chaya$^{1,^*}$, Ate Poortinga$^{2,^{**}}$, Keo Nimol$^{1}$,  Se Sokleap$^{1\dagger}$, Mon Sophorn$^{1\dagger}$, Phy Chhin$^{1\dagger}$\\
    \textbf{Andrea McMahon}$^{2}$, \textbf{Andrea Puzzi Nicolau}$^{2}$, \textbf{Karis Tenneson}$^{2}$, \textbf{David Saah}$^{2,3}$ \\
    $^1$Department of Agricultural Land Resources Management,  General Directorate of Agriculture,\\ Ministry of Agriculture, Forestry and Fisheries, Cambodia \\ $^2$ Spatial Informatics Group, LLC, 2529  Yolanda Ct., Pleasanton, CA 94566, USA \\ $^3$Geospatial Analysis Lab, University of San Francisco,\\ 2130 Fulton St., San Francisco, CA 94117, USA\\
    $^*$\texttt{chaya.veasna@gmail.com}, $^{**}$\texttt{apoortinga@sig-gis.com}
}

\renewcommand{\shorttitle}{Cashew Expansion in Cambodia}

\hypersetup{
pdftitle={Is Cambodia the World's Largest Cashew Producer?},
pdfsubject={q-bio.NC, q-bio.QM},
pdfauthor={David S.~Hippocampus, Elias D.~Striatum},
pdfkeywords={First keyword, Second keyword, More},
}

\begin{document}
\maketitle

\begin{abstract}
Cambodia’s agricultural landscape is rapidly transforming, particularly in the cashew sector. Despite the country's rapid emergence and ambition to become the largest cashew producer, comprehensive data on plantation areas and the environmental impacts of this expansion are lacking. This study addresses the gap in detailed land use data for cashew plantations in Cambodia and assesses the implications of agricultural advancements. We collected over 80,000 training polygons across Cambodia to train a convolutional neural network using high-resolution optical and SAR satellite data for precise cashew plantation mapping. Our findings indicate that Cambodia ranks among the top five in terms of cultivated area and the top three in global cashew production, driven by high yields. Significant cultivated areas are located in Kampong Thom, Kratie, and Ratanak Kiri provinces. Balancing rapid agricultural expansion with environmental stewardship, particularly forest conservation, is crucial. Cambodia's cashew production is poised for further growth, driven by high-yielding trees and premium nuts. However, sustainable expansion requires integrating agricultural practices with economic and environmental strategies to enhance local value and protect forested areas. Advanced mapping technologies offer comprehensive tools to support these objectives and ensure the sustainable development of Cambodia's cashew industry.
\end{abstract}


\section{Introduction}
Cambodia has undergone considerable land cover changes in recent years, influenced by economic development, agricultural expansion, infrastructure projects, and population dynamics. The Cambodian Government’s introduction of large-scale Economic Land Concessions (ELCs) in 2005 facilitated a marked increase in agricultural activities, plantation developments, and other land use transformations. In 2012, Directive 01 was introduced to speed up a land titling program, further influencing land use patterns. This period has also seen a notable rise in agricultural productivity, which has been instrumental in enhancing the economic status of the people in Cambodia. However, these changes, while contributing to the country’s economic landscape, have also led to alterations in land use patterns, affecting forest areas and biodiversity \citep{teck2023land}.

Cashews can grow on a range of soils but are best suited to well-drained sandy loam soil. Cashews are often grown on marginal soils where other fruit and nut trees cannot grow. They require little rainfall (1000mm p.a. minimum) \citep{Fitzpatrick2019Cashew}. For Cambodian cashews, little value is lost during processing as they are easy to shell while they are cultivated using low amounts of pesticides. The government's strategic focus on cashews includes cultivation, refining, and export. Their vision is to develop production and processing and organize a competitive and sustainable product. An important aspect to highlight is value addition. Cambodia’s ambition is not only to produce and export raw cashews but also to develop a domestic cashew industry with significant value addition. Reducing reliance on external processing and developing a local processing sector presents a significant opportunity for Cambodia. Currently, only a small portion of the total cashew production is processed locally, whereas the majority is processed and packed in Vietnam, from where it is sold to other global markets \citep{CambodiaCashew2023}. A recent study of \cite{JICAReport2024} indicates that the majority (80\%) of cashew farmers are small-scale, owning less than 5 hectares of land, followed by medium-scale farmers with 5-10 hectares (16\%), and a minority (6\%) owning more than 10 hectares. Despite Vietnam being the top global exporter of cashews, it imports 65\% of its raw materials from Cambodia, which in turn exports over 90\% of cashews to Vietnam. An evaluation by the International Finance Corporation (IFC) and European Union (EU) highlights that Cambodian cashews, constituting 24-28\% kernel yield from raw cashew nuts, are comparable to Vietnam's in yield but are larger, more valuable, easier to shell, and fetch premium prices due to reduced pesticide use \cite{WorldBank2010}.

Accurate crop mapping plays a crucial role in facilitating evidence-based decision-making within the agricultural sector. By providing precise information on the spatial distribution and extent of crops, decision-makers can optimize resource allocation, enhance agricultural practices, and support sustainable development goals. Policy implications of accurate crop mapping include targeted interventions for specific regions or crops, improved market analysis, and the identification of areas requiring agricultural development. Accurate crop mapping informs sustainable land use planning and conservation efforts. It enables policy makers to identify areas at risk of landcover conversions and implement measures for preserving valuable forest ecosystems. This contributes to the National Agriculture Development Policy 2022-2030 growth of competitive agriculture value chains that provide high-quality, safe, and nutritious products while increasing sustainable land management.
Accurate crop mapping contributes to monitoring greenhouse gas emissions and establishing baselines for crop finance. This information could enable the implementation of climate finance initiatives and the development of strategies to reduce agricultural emissions, promoting sustainable and environmentally friendly agricultural practices. This aligns with the national long-term strategy for carbon neutrality, helping Cambodia to achieve the Paris Agreement commitment to limit global warming to 1.5 C. Improved crop mapping also assists with reporting non-CO2 gases emitted from agriculture land use practices, as recommended by the UNFCCC technical assessment report on the second forest reference level of Cambodia submitted in 2021.

Accurately mapping cashew plantations is also critical for adhering to global compliance frameworks such as REDD+ under the UNFCCC and the new EU regulations mandating deforestation-free commodities. While participation in REDD+ and adherence to the EUDR is voluntary, countries choose to comply to access specific markets and benefits. These initiatives require detailed land use data to effectively track agricultural expansion into forested areas. Yet, there is a notable scarcity of precise maps and data regarding the total area of cashew plantations and their environmental impact. Current estimates often rely on data from regional governments and the cashew processing industry, which tend to be aggregated at the level of administrative units, lacking the necessary spatial resolution and scientific scrutiny. While sample-based methods are regarded as the scientific gold standard for data accuracy, they demand extensive sample sizes and can be time-consuming, especially when applied randomly. In the case of multiple land use and land use change classes, stratified sampling, enhanced by land cover maps, offers a more focused approach to data collection. However, the absence of high-quality cashew plantation maps presents a significant challenge to this method.

Land cover maps often contain generalised information on crops or tree plantations and don't contain labels on specific agricultural commodities. The relatively coarse resolution of freely available satellite imagery and similarity of spectral signatures, e.g., cashew to other tree crops like rambutan and mango, made it difficult to distinguish between those classes. However, higher-resolution satellite data and the application of advanced deep learning and computer vision technologies has proven to be promising in increasing the accuracy. Recent studies of, e.g., \citet{kalischek2023cocoa,poortinga2021mapping,masolele2024mapping} show that convolutional neural networks \citep{ronneberger2015u} outperform traditional machine learning methods and provide highly accurate maps for specific land cover types. However, training these models requires a large amount of spatial explicit training data \citep{parekh2021automatic,mayer2021deep}.

In this study, we collected extensive land cover data for Cambodia. These data were used to train a convolutional neural network using PlanetScope, Sentinel-1, Sentinel-2, and Landsat imagery. The model was applied to the whole of Cambodia to identify cashew probability. The maps were then combined with forest loss information to estimate the relative contribution of (historical) forest loss to cashew planting as well as stratification for area estimations. 

\section{Methods}\label{sec11}

\subsection{Training data}

Training data was collected through field surveys, drone surveillance, and satellite imagery interpretation across 19 out of 24 provinces in Cambodia. For each targeted province, at least six 4 by 4 km squares were designated in specific agricultural areas of interest, focused around areas with recent forest to agriculture transition. For provinces with a high concentration of Cashew, a larger number of squares were designed. These boxes were all hand digitised, wall-to-wall, resulting in a comprehensive dataset of 83,034 polygons. The dataset contains a large number of labels, with 21 of the majority categories shown in Table \ref{table:land_cover}. The "Other" category contains the labels on specific forest types, sparse crops and objects like farms and pegodas. For each polygon a label was added for the Google Earth's base layer, drone imagery, and PlanetScope satellite images.  A total of 10,147 polygons were classified as cashew plantations. For cashews, also the age of the trees was included as a separate label. For young cashew trees, the area is bare, covered with grass and shrubs, or inter-cropped with annual crops, predominately cassava.

\begin{table}[ht]
\centering
\caption{Distribution of Land Cover Categories}
\label{table:land_cover}
\begin{tabular}{lr}
\hline
\textbf{Category} & \textbf{Polygon Count} \\
\hline
Banana & 314 \\
Bare land & 13921 \\
Cashew & 10147 \\
Cassava & 2939 \\
Coconut & 295 \\
Durian & 241 \\
Forest Cover & 776 \\
Grassland & 30341 \\
House & 1157 \\
Longan & 727 \\
Maize & 175 \\
Mango & 2644 \\
Orange & 119 \\
Other & 463 \\
Pepper & 695 \\
Rice & 307 \\
Road & 646 \\
Rubber & 2516 \\
Shrubland & 6754 \\
Village & 3296 \\
Water Body & 4561 \\
\hline
\textbf{Total} & \textbf{83034} \\
\hline
\end{tabular}
\end{table}

\begin{figure}[ht]
    \centering
    \includegraphics[width=\linewidth]{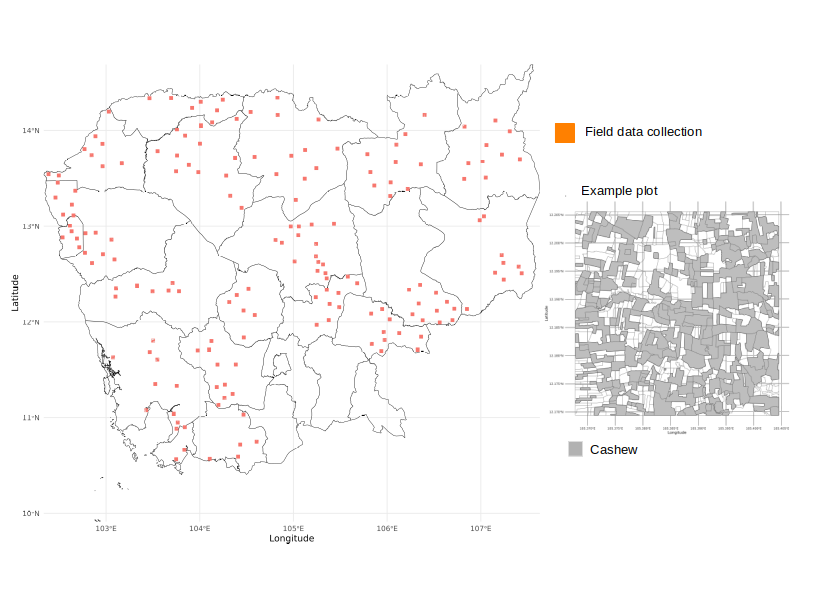}
    \caption{Spatial distribution of the all the different categories of training data.}
    \label{fig:field}
\end{figure}

\subsection{Neural Network model}

The architecture of the neural network is based on an attention-enhanced U-Net \citep{oktay2018attention}, which incorporates four max-pooling layers to facilitate feature reduction and computational efficiency. The integration of attention mechanisms occurs within the skip connections, bridging the encoder and decoder segments, thereby enhancing feature relevance during the reconstruction phase. The network is designed to ingest four input channels for varying resolutions in the satellite imagery sources. At the highest level, PlanetScope imagery, characterised by a 256x256 pixel patch with a 5-meter spatial resolution, is fed into the network. Subsequent layers integrate Sentinel-1 and Sentinel-2 imagery at a reduced resolution of 128x128 pixels following the first max-pooling operation, catering to their 10-meter spatial resolution. Further refinement through max-pooling allows for incorporating coarser resolution bands from Sentinel-2, including NIR, SWIR, and red-edge bands at a 20-meter resolution, represented as 64x64 pixel patches. The integration process ends with adding Landsat imagery, which, due to its 40-meter resolution, is inputted as 32x32 pixel patches. As such, Landsat was sampled at a coarser resolution than the native resolution to match the network's architecture.  This hierarchical incorporation of multispectral data, combined via concatenation at various stages of the network, ensures a comprehensive feature representation. In the decoding phase, all inputs are progressively upsampled back to the original 256x256 pixel resolution. 

The U-Net architecture was chosen as it combines low-level details from initial layers with high-level features from deeper layers through skip connections. Integrating varied satellite data, from high-resolution PlanetScope to multispectral Sentinel and Landsat, provides a comprehensive spectrum of information. This approach enables the model to identify subtle land cover differences while improving accuracy and adaptability across the landscape. Sequentially integrating satellite products with decreasing resolution mimics a hierarchical learning process. The model starts with high-resolution details and progressively incorporates broader contextual information from coarser-resolution images. Combining data from various sources makes the model more robust to variations in any single data source, such as differences in lighting conditions, cloud cover, or seasonal changes. It also helps the model generalize better across different geographic regions and conditions, as it learns to leverage complementary information from multiple datasets. 

One key advantage of employing georeferenced data labels in this study is their suitability for use with satellite images from various dates, assuming the land cover categories remain unchanged over time. To leverage this, we selected the highest quality optical satellite imagery, carefully chosen to exclude images with clouds, cloud shadows, and other visual imperfections. These selected images were then randomly integrated into the training patches. This method was deliberately chosen to equip the model with the ability to generalize across a wide range of image conditions and combinations. For the Sentinel-1 data, we utilised yearly averaged data.

\begin{figure}[ht]
    \centering
    \includegraphics[width=\linewidth]{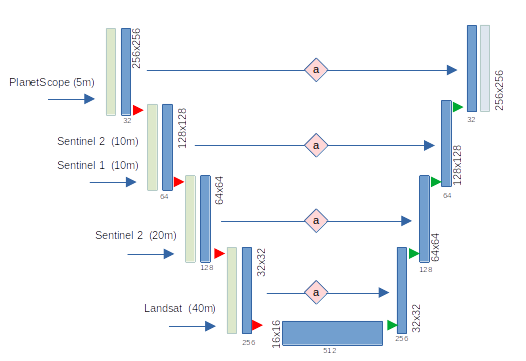}
    \caption{The U-net attention architecture utilizing data from PlanetScope, Sentinel-2, Sentinel-1, and Landsat. PlanetScope data is integrated at the highest level of the network, while Sentinel-2 and Sentinel-1 bands are incorporated post-maxpooling operations.}
    \label{fig:network}
\end{figure}

\subsection{Model Training}

The training of the model was conducted in two distinct phases to address the challenges posed by class imbalances and train a generalised model that recognizes the landscape in Cambodia. Initially, the model was trained using 43 labels with a softmax activation function paired with a dice coefficient function. We generated a large number of training patches over the training samples. These patches were produced by random sampling, ensuring overlap among them to enhance robustness. Additionally, data augmentation techniques such as flipping and rotating were applied to these patches to further increase the diversity of training data. This resulted in around 250,000 training patches for training and another 60,000 for validation with a batch size of 16. Early stopping was applied to prevent overfitting. In the subsequent training phase, the model's architecture was retained, but the encoder was frozen to refine its focus on cashew classification as binary data. Training patches were generated in a similar fashion, but focused on data patches that contain cashew to account for data imbalances. This resulted in around 60,000 patches for training and 15,000 for validation. It should be noted that only cashew polygons with an age greater than 3 years were included. During this phase, a sigmoid activation function was employed alongside a boundary loss function \citep{wang2022active}. The choice of boundary loss was strategic, designed to impose higher penalties for misclassifications that occur further from the actual boundaries of cashew fields. This approach was particularly useful in mitigating the effects of potential misalignments between the satellite-derived labels and the actual ground-truth locations of the cashew fields. Table \ref{tab:validation_metrics} shows the performance indices for the first and second training phases. It can be seen that the overall performance of the model using all labels is quite good, with an accuracy of 0.85. The performance of the binary cashew model is higher, with an accuracy of 0.97.

\begin{table}[ht]
\centering
\caption{Validation scores model training}
\label{tab:validation_metrics}
\begin{tabular}{|l|l|l|l|l|}
\hline
\textbf{Category} & \textbf{Accuracy} & \textbf{Precision} & \textbf{Recall} & \textbf{F1 Score} \\ \hline
All Labels        & 0.8413            & 0.8718             & 0.8131          & 0.8412            \\ \hline
Cashew            & 0.9744            & 0.9752             & 0.9735          & 0.9743            \\ \hline
\end{tabular}
\end{table}

\subsection{Inference}

Satellite data covering Cambodia were gathered, prioritizing the highest quality datasets (no clouds and cloud shadows) for the year 2023. This data was subsequently processed by the neural network to compute probabilities at the pixel level. The model underwent two distinct runs: initially with fixed weights to establish a baseline, followed by a second run where a dropout rate was introduced during inference to accommodate potential variability in the data. With dropout activated, the model produced 10 distinct realizations, the average of which was then computed. These averaged results, combined with the outcomes from the initial run, constituted the final output, which was subsequently utilised for the model's validation process.

\subsection{Validation and area estimation}

Data validation was conducted at the provincial level to ensure the accuracy and reliability of the cashew distribution map. To mitigate the impact of network artifacts, especially in regions devoid of training data, the cashew map underwent a cleaning process before integration with the SERVIR landcover map \citep{saah2019land,saah2020primitives,poortinga2019mapping}. For a more precise analysis, the probability map associated with the cashew distribution was segmented into seven distinct categories, ranging from 30\% to 100\% probabilities. This stratification facilitated a better understanding of the model's performance across various probability thresholds and provinces while enabling more precise area estimates. This stratification approach is particularly advantageous for lower probability categories, where increased sampling density relative to the size of each stratum enhances the accuracy and area estimations. For the cashew strata, a higher sample-to-area ratio was used. Moreover, the SERVIR landcover map was used for a strategic reduction in sampling density within categories anticipated to exhibit minimal confusion, such as water bodies and dense forests. The samples were analysed in collect earth \citep{saah2019collect} by image interpretation using Planetscope imagery and the high resolution base layers. The confusion matrix and associated areas were calculated using the methods formulated by \citep{olofsson2014good,olofsson2013making,olofsson2020mitigating}.

\subsection{Forest loss dataset}

The forest loss data was generated using the Global Land Analysis and Discovery (GLAD) system. Contrary to the global data \citep{hansen2013high}, our algorithm is specifically tailored for regional applications \citep{potapov2019annual}. GLAD leverages Landsat time-series data, complemented by Lidar-based vegetation structure prediction models, to accurately map woody vegetation canopy cover and height. The GLAD system was used to generate data on Tree Canopy Cover (TCC), Tree Canopy Height (TCH), and forest loss using the Landsat collection-2 data for the period 2000 - 2022. In this study we use the forest loss dataset, which contains binary information on forest loss for each year. \citet{potapov2019annual} reported an accuracy of 0.93 (SE 0.005) for the GLAD dataset in detecting tree cover loss, based on a validation analysis of 2,964 Landsat pixels. They further reported a user's accuracy of 0.84 (SE 0.02) and a producer's accuracy of 0.75 (SE 0.02) for the loss class. This loss data was used to calculate the year of the earliest forest loss of the cashew area. 

\section{Results}\label{sec2}

\begin{figure}[ht]
    \centering
    \includegraphics[width=\linewidth]{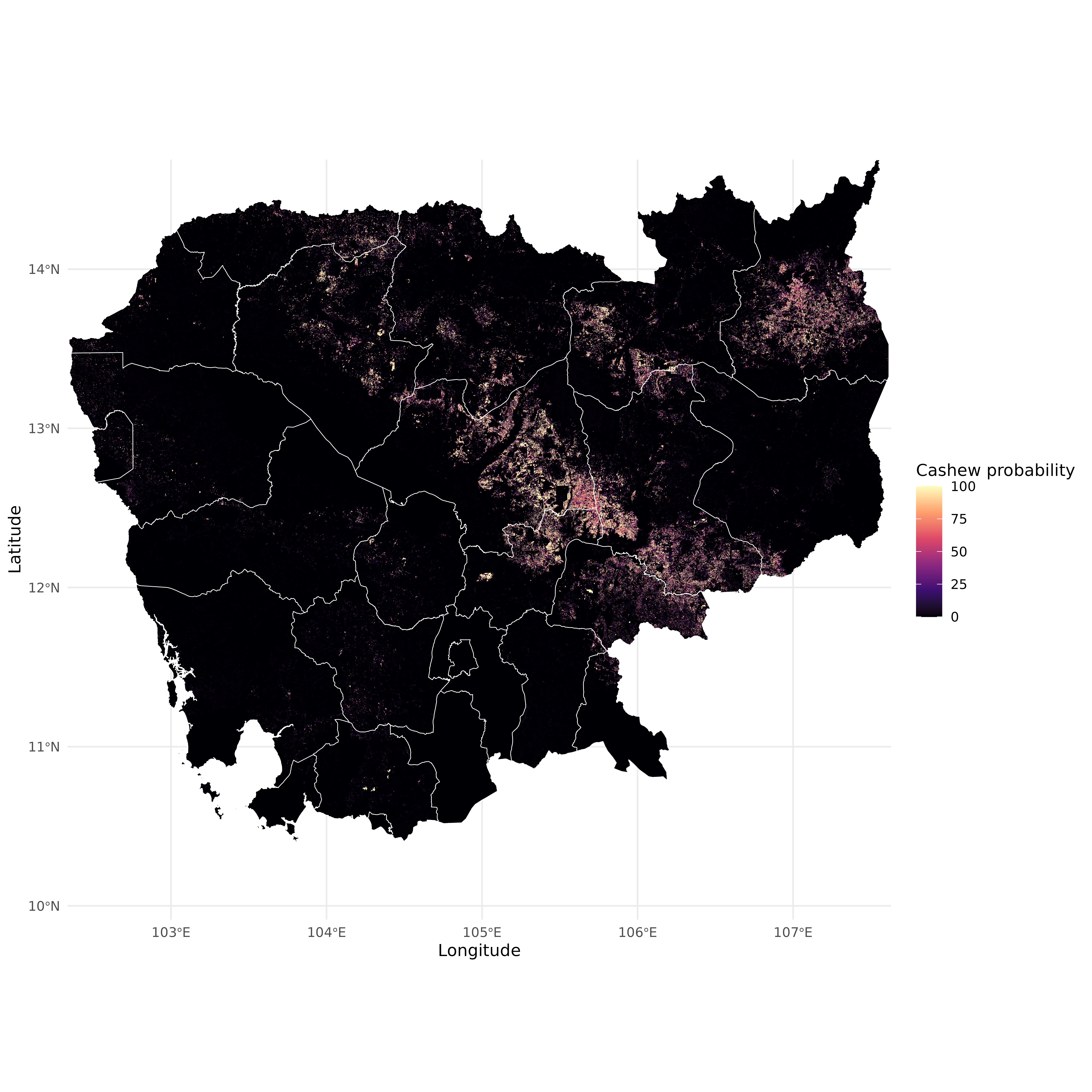}
    \caption{The 2023 cashew probability map for Cambodia.}
    \label{fig:cashew}
\end{figure}

The distribution of cashew plantations, as shown in Figure \ref{fig:cashew}, exhibits a clear geographical pattern within Cambodia. Notably, the provinces of Kampong Thom, Kratie, and Ratanak Kiri are identified as having the densest cashew populations. In contrast, the concentrations of cashew trees are  lower in the country's western and southwestern regions. The clear demarcation between high-probability cashew pixels and those with low probability indicates the model's efficacy in discerning cashew presence across the landscape. The probability distribution primarily clusters at the extremes, with a significant majority of pixels either in the 90-100\% probability bracket, indicating high confidence in cashew identification, or within the 0-10\% range for areas with no cashew.

Table \ref{tab:province_data} shows the accuracy scores, the unbiased area estimates with their confidence interval (CI), the yield, and production with their 95\% confidence level. Total production was calculated from the area and yield estimates, where the latter were obtained from the General Directorate Agriculture (GDA) and not measured or independently verified. It can be seen that Kampong Thom has the highest cashew area, followed by Kratie and Ratanak Kiri. For the largest producing provinces, we obtained the highest accuracies varying between 93.6\% for Kratie and 94.9\% for Ratankiri. For Stung Treng, Oddar Meanchey and Siemreap, also large Cashew producers, we also obtained high accuracies above 90\%. For many of the smaller cashew producing areas, accuracies were lower. The 95\% confidence intervals for the area estimates are all within acceptable ranges, with 15\% of the total estimate for Koh Kong and 3\% for Mondul Kiri. Our total area estimate for the whole country is 580,117 \textpm 27,710 ha. Translating these areas into yields gives a maximum total production of 816,459 \textpm 37,139 ha, but this is lower in reality as not all fields are fully productive all the time.

\begin{table}[h!]
\centering
\caption{The accuracy, the area cultivated, the confidence interval (CI), the average yield per hectare, and the total agricultural production (in tons) for each province. The reported production assumes all fields are fully productive, which does not reflect actual conditions. For provinces with small cashew areas, the $^{*}$ indicates data was adopted from the provincial department of agriculture forestry and fisheries (PDAFF).}
\label{tab:province_data}
\begin{tabular}{|l|r|r|r|r|}
\hline
\textbf{Province}       & \textbf{Accuracy (\%)} & \textbf{Area (ha)} & \textbf{Yield (ton/ha)} & \textbf{Production (ton) [CI (ton)]} \\ \hline
Banteay Meanchey        & 89.1                   & 2,137              & 1.60                    & 3,419 [214]                \\ \hline
Battambang              & 75.7                   & 3,139              & 1.25                    & 3,924 [419]                \\ \hline
Kampong Cham            & 89.6                   & 46,582             & 1.30                    & 60,557 [3,661]             \\ \hline
Kampong Chhnang         & 82.2                   & 1,687              & 1.50                    & 2,531 [275]                \\ \hline
Kampong Speu            & 84.3                   & 2,616              & 1.50                    & 3,924 [498]                \\ \hline
Kampong Thom            & 94.1                   & 147,703            & 1.49                    & 220,077 [9,883]            \\ \hline
Kampot                  & 73.9                   & 496                & 1.50                    & 744 [69]                   \\ \hline
Kandal                  & -                      & -                  & -                       & -                          \\ \hline
Koh Kong                & 68.7                   & 1,024              & 1.37                    & 1,403 [212]                \\ \hline
Kratie                  & 93.6                   & 102,520            & 2.00                    & 205,039 [4,270]            \\ \hline
Mondul Kiri             & 89.0                   & 9,858              & 1.00                    & 9,858 [303]                \\ \hline
Phnom Penh              & -                      & -                  & -                       & -                          \\ \hline
Preah Vihear            & 85.2                   & 28,965             & 1.40                    & 40,551 [2,331]             \\ \hline
Prey Veng               & 85.7                   & 1,119              & 1.20                    & 1,343 [114]                \\ \hline
Pursat                  & 77.7                   & 1,704              & 1.25                    & 2,130 [259]                \\ \hline
Ratanak Kiri            & 94.9                   & 97,258             & 0.68                    & 66,135 [3,068]             \\ \hline
Siemreap                & 92.5                   & 35,914             & 1.75                    & 62,850 [2,314]             \\ \hline
Sihanoukville$^{*}$     & -                      & 542                & 1.40                    & 759 [-]                    \\ \hline
Stung Treng             & 91.2                   & 44,250             & 1.20                    & 53,100 [3,763]             \\ \hline
Svay Rieng              & 82.8                   & 1,532              & 1.70                    & 2,604 [223]                \\ \hline
Takeo$^{*}$             & -                      & 223                & 1.50                    & 335 [-]                    \\ \hline
Otdar Meanchey          & 92.9                   & 13,818             & 1.50                    & 20,727 [659]               \\ \hline
Kep$^{*}$               & -                      & 11                 & 1.20                    & 13 [-]                     \\ \hline
Pailin                  & 77.3                   & 616                & 1.50                    & 924 [125]                  \\ \hline
Tboung Khmum            & 91.3                   & 36,403             & 1.47                    & 53,512 [4,481]             \\ \hline
\textbf{Total}          &                        & \textbf{580,117}   & -                       & \textbf{816,459 [37,139]}  \\ \hline
\end{tabular}
\end{table}

\begin{figure}[ht]
    \centering
    \includegraphics[width=0.8\linewidth]{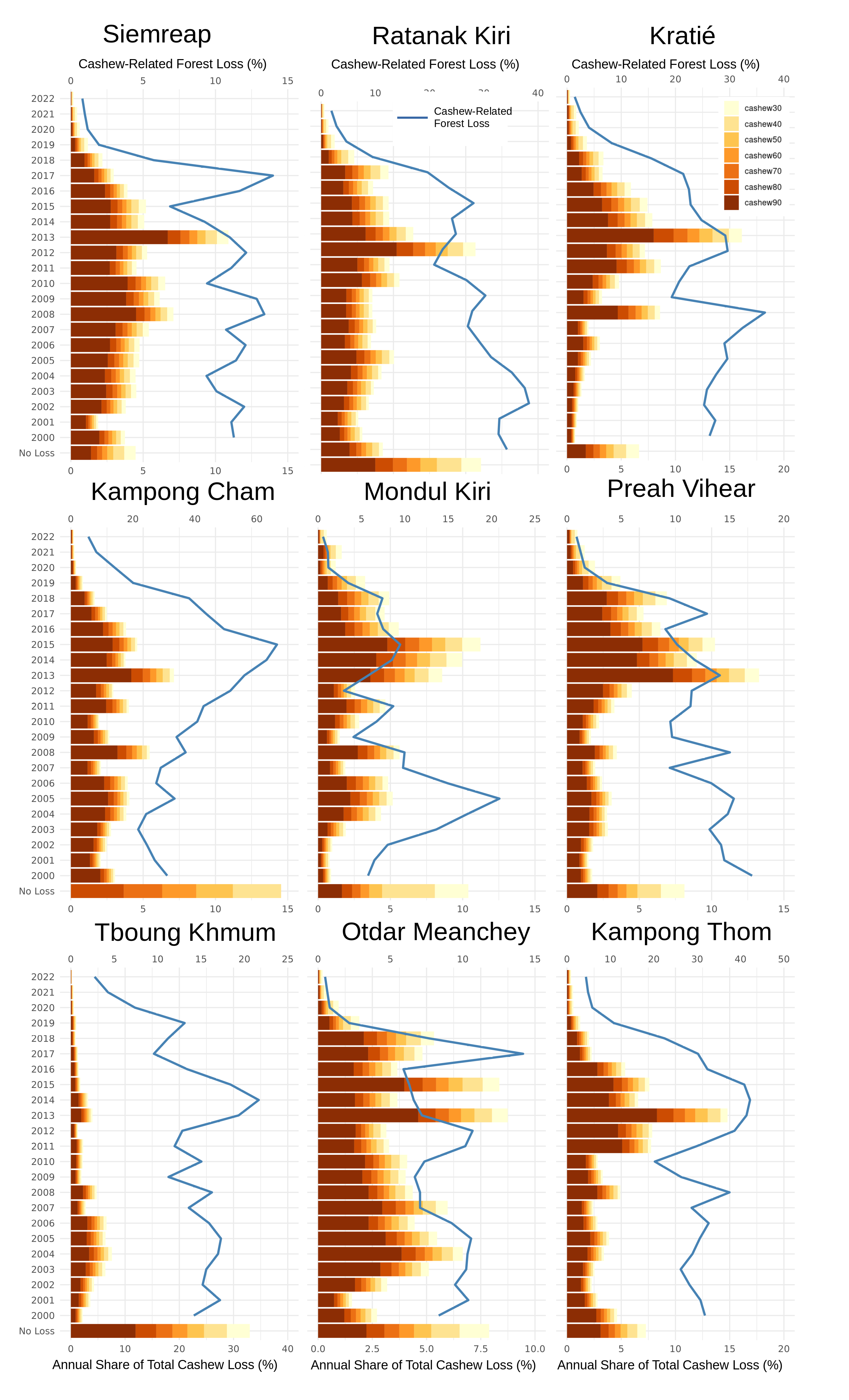}
    \caption{Forest loss per cashew probability category annually, normalised to 100\%; the blue line (top axis) shows cashew-related forest loss as a percentage of total yearly loss from 2000 to 2022.}
    \label{fig:bar}
\end{figure}

The cashew probability map was combined with the GLAD yearly forest loss data (Figure \ref{fig:bar}). The bars in the graph represent the proportion of forest loss within each cashew probability category annually, with the cumulative loss for all categories in a given year equaling 100\%. This analysis utilizes data from the first recorded loss year, covering the period from 2000 to 2022. Cashew fields that remained unaffected by forest loss during this time-frame are denoted as 'no loss'. The top axis, accompanied by a blue line, illustrates the percentage of forest loss attributable to cashew cultivation relative to the total annual forest loss.

It can be seen in Figure \ref{fig:bar} that there is a noticeable acceleration in cashew-related forest loss from 2012 in all provinces with a gradual decline from 2015. For Tboung Khmum, we found the highest percentage of areas where cashew cultivation has not led to forest replacement over the past 22 years. Kampong Cham, Ratana Kiri, and Mondulkiri closely follow this. Conversely, Kampong Thom, Kratie, and Siemreap are characterised by lower rates of non-forest loss cashew cultivation. In Kampong Cham, despite a significant number of cashew plantations not contributing to landcover change, cashew cultivation emerged as a main cause of the province's forest loss, accounting for 60\% of the total in 2013. Similarly, in Kampong Thom, Kratie, and Ratana Kiri, cashew cultivation has played a notable role in landcover conversions. Conversely, while many of Mondulkiri's current cashew plantations were established on land previously covered by forest, they do not presently lead the causes of forest loss. In Preah Vihear, the impact of cashew cultivation on forest loss has been gradually diminishing, whereas in Tboung Khmum, the contribution has remained consistent, fluctuating between 10 and 20\%. The model can not detect young cashew plantations from space, as the fields often look bare or cashew is inter-cropped with cassava. This has caused a sharp decline in cashew-related forest loss in recent years.

\begin{figure}[ht]
    \centering
    \includegraphics[width=\linewidth]{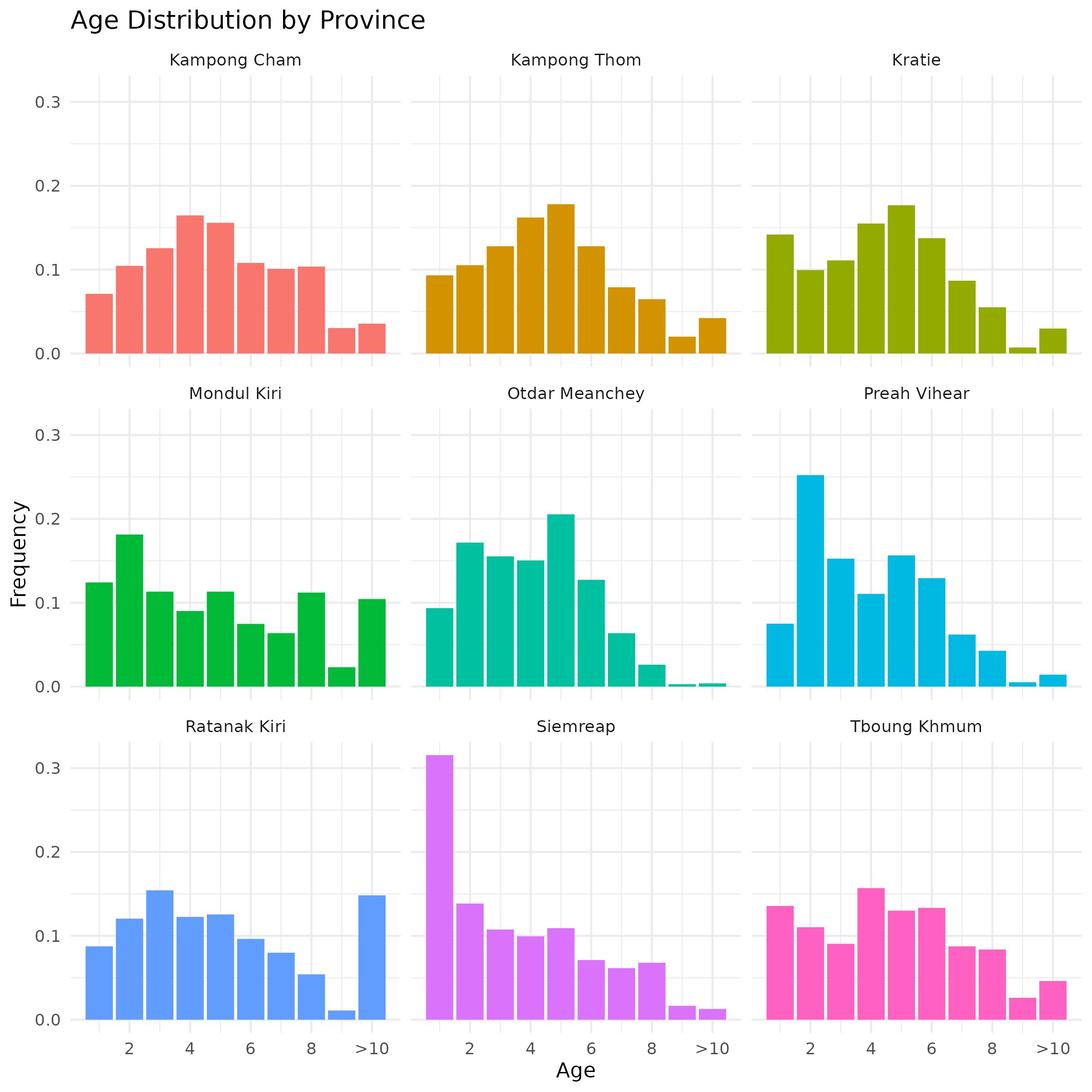}
    \caption{Cashew age distribution in main cashew producing areas.}
    \label{fig:age}
\end{figure}

As young cashew fields are difficult to detect, we analysed the age distribution in the training data for the different provinces (figure \ref{fig:age}). The age distribution can help understand cashew expansion in the near future as the cashew signal can be picked up from space after the cashew trees reach a maturity of about 3 years. Cashew trees typically have a productive lifespan of around 25 to 30 years, after which they need to be replaced. This periodic renewal is necessary to maintain optimal yield and manage pest infestations and diseases. Additionally, farmers may clear cashew plantations to plant more profitable crops. It is noticeable that most older plantations can be found in Ratanakiri, the youngest ones in Siem Reap. For Tboung Khum, we found the highest number of plantations around 2 years. For the major cashew producing regions Kampong Thum and Kratie, we find a large portion of the plantations between 4 and 6 years old.

\section{Discussion}\label{sec12}

Deep learning technologies have greatly advanced our capabilities for mapping specific crop commodities such as cashews. However, models are very data-hungry, and we found a performance degradation in regions further away from the major cashew-producing areas. This can be seen in the reported accuracies, where performance for the major producing cashew provinces is high, but much lower for provinces with a small area. For those provinces, the model shows confusion between other tree crops, cropland, and shrublands. It should also be noted that cashew fields become identifiable from space only after 3-5 years, and when the trees are small they are often inter-cropped predominantly with cassava. This phenomenon is also the main reason for a sharp drop in cashew-related forest loss, as found in figure \ref{fig:bar}. Furthermore, it is noted that the age distribution of cashews presented was derived from the training sample, which was collected at specific areas of interest in the field. As such, they do not represent a statistically randomized sample across each province. We did not include replacement in the analysis as most cashew plantations are of recent date with no history of previous cashew cultivation.    

Our total area estimate for cashews in Cambodia is 580,117 ha, which is likely to increase as there are still many young cashew plantations. Cashew plantations younger than 2 years are difficult to detect using satellite imagery as these fields look bare from the satellite or they are inter-cropped with cassava. Although cashew is not currently under the EUDR, the 2020 cut-off date is relevant for understanding the land use history and changes before and after 2020. We can anticipate a more accurate estimate in 2024 and 2025 as the young trees become more easily detectable from satellite imagery. Our estimate is lower than the number reported by the Cashew Nut Association of Cambodia (CAC), which estimates that this year's cashew plantations covered approximately 700,000 hectares \cite{Pisei2023CambodiaCashew}. Table \ref{tab:cashew_production} shows the total cashew area for the major cashew-producing countries worldwide. Côte d'Ivoire has the largest area, almost totaling 2 million hectares, followed by India (1.16 million ha) and the United Republic of Tanzania (712,624 ha). This would make Cambodia the fourth-largest producer by area, still larger than the 519,000 that was reported by \citet{yin2023mapping} in a recent study. However, Cambodia's reported yields are much higher than those of Indonesia, India, and cashew producers on other continents. Our total production estimate of 903,598 tons is higher than the reported production in Côte d'Ivoire, the number one producer, but might be at the higher end as not all fields are productive all the time. The CAC claims Cambodia was the largest producer of cashews in 2021, with a total of 1.18 million tons \cite{Pisei2023CambodiaCashew}. While our yield and production have higher uncertainties compared to our area estimates, as these cannot be easily estimated from space, it is clear that Cambodia is among the top 3 producers in the world.

\begin{table}[h!]
\centering
\caption{Cashew Production Data for Major Producing Countries (2021) from FAOSTAT \citep{FAOSTAT2024}}
\label{tab:cashew_production}
\begin{tabular}{|l|r|r|r|}
\hline
\textbf{Country or Area} & \textbf{Area (ha)} & \textbf{Yield (tons/ha)} & \textbf{Production (tons)} \\ \hline
Vietnam                 & 294,901            & 1.354                    & 399,296                    \\ \hline
Benin                    & 406,893            & 0.370                    & 150,428                    \\ \hline
Brazil                   & 427,144            & 0.260                    & 111,100                    \\ \hline
Indonesia                & 477,977            & 0.357                    & 170,447                    \\ \hline
United Republic of Tanzania & 712,624          & 0.296                   & 210,794                    \\ \hline
India                    & 1,159,000          & 0.637                    & 738,051                    \\ \hline
Côte d'Ivoire            & 1,989,861          & 0.421                    & 837,930                    \\ \hline
\end{tabular}
\end{table}

Our analysis indicates a discernible decline in forested areas within Cambodia, corroborated by governmental reports. The expansion of cashew cultivation has emerged as a significant contributor to the reduction of forest cover in certain provinces. This agricultural shift aligns with broader economic strategies that have been important in transforming land use patterns. Notably, such policy implementations in Cambodia have been more recent compared to other nations, where similar land use changes occurred before the era of high-resolution satellite monitoring. It is imperative to contextualize these changes within Cambodia’s historical backdrop, characterised by socio-economic upheaval and the pursuit of stability and growth. The push towards cashew production has been part of a larger effort to enhance food security and foster economic development. Over the recent period, there has been a marked improvement in the nation’s economic landscape, with the cashew industry playing a key role in augmenting the export market. Given the global demand for cashews, driven by their recognised health benefits and nutritional value. 

While forest replacement by cashews is a major concern, cashew cultivation is generally considered low-impact. They involve small-scale farming practices, minimal water use, and no irrigation. They are often grown on marginal land and have the capacity to grow deep roots to access the water table. Planted from seeds or seedlings usually locally produced, the planting of cashew trees does not involve large-scale soil disturbance or the use of machines. Once planted, cashew trees have a long, productive life. A study of \citet{noiha2017floristic} indicates that cashews are high-capacity carbon sinks. However, in Cambodia, cashew production has expanded by extension rather than yield improvement. Projects aimed at improved harvest and post-harvest management and pest control can further help improve production \citep{Fitzpatrick2019Cashew}.

In the updated Nationally Determined Contribution (NDC), Cambodia has pledged to continue its collaborative efforts with the global community to reduce greenhouse gas (GHG) emissions and enhance the nation's adaptive capacity, directing its resources towards climate-resilient and sustainable development. This commitment extends to ensuring that cashew production is sustainable and does not encroach on forested areas, aligning with Cambodia's national strategy for reducing deforestation and forest degradation. Additionally, Cambodia aims to develop transparent supply chains, potentially opening access to high-value markets under frameworks like the EU's compliance system for deforestation-free commodities. The methodologies employed in this study offer a comprehensive analysis of the cashew production in Cambodia, in an effort to inform a cross-sectoral action for reaching Cambodia's climate change goals vis-a-vis  economic development objectives. 

\section{Conclusion}

Cambodia now ranks among the top three global cashew producers, along with India and Côte d'Ivoire, boasting an estimated 580,000 hectares of cultivation, particularly in Kampong Thum, Kratie, and Ratanak Kiri. The country's ascent in cashew production is attributed to its high-yielding trees and premium nuts. For sustained growth, it's crucial to improve current agricultural practices with economic strategies to enhance local value while also adhering to environmental regulations to prevent expansion into protected areas. The observation of numerous young cashew fields suggests an expected increase in cultivation areas, promising further expansion in Cambodia's cashew production. Further investment in national expertise for land cover and crop mapping is important prerequisite to informed crop production, and sustainable land use planning. 

\section{Acknowledgements}

This project was supported through a 3-year partnership between the General Directorate of Agriculture, the U.S. Government program SilvaCarbon and the Food and Agriculture Organization of the United Nations (FAO) focused on improving the mapping processes needed to achieve timely updates for the enhanced transparency framework under the Paris Agreement. We would like to extend our gratitude to Marija Kono, from SilvaCarbon in the United States for her support during this project. We also extend our thanks to Jamil Mahmood from UNDP Cambodia, Sophyra Sar and Mathieu Van Rijn from FAO and Dr. Seng Vang from Department of Agricultural Land Resources Management in Cambodia for their contributions and guidance. Their expertise and assistance have been invaluable. We thank Vanna Teck for support collecting validation samples. The partnership also benefited from the expertise of different technical officers at the Ministry of Environment, who were involved in the different steps.

\section{Data and Model Availability}
The code for our model and data generation scripts is openly accessible on our GitHub page: \href{https://github.com/SERVIRSEA/cashew}{SERVIR-SEA Cashew Project}. Model weights are available upon request. The training data were cannot be shared publicly.
 
 The map related to our project can be viewed at the following link: \href{https://servir-ee.projects.earthengine.app/view/cambodiacashew}{Cambodia Cashew Map}.

\end{document}